\documentclass[aps,twocolumn]{revtex4}
\usepackage[utf8]{inputenc}
 \usepackage{amsmath}
\usepackage{amssymb}
\usepackage{latexsym}
\usepackage{epsfig}
\usepackage{color}

\usepackage{amsmath,amssymb}
\usepackage{graphicx} 
\usepackage{gastex}
\usepackage{epstopdf}
\usepackage[colorlinks=true,citecolor=blue,linkcolor=blue,urlcolor=black]{hyperref}
\usepackage{color}

\begin{document}
\title{Modified cosmology through Barrow entropy}
\author{Ahmad Sheykhi\footnote{asheykhi@shirazu.ac.ir}}
\address{Department of Physics, College of
Sciences, Shiraz University, Shiraz 71454, Iran\\
Biruni Observatory, College of Sciences, Shiraz University, Shiraz
71454, Iran}

 \begin{abstract}
We investigate the cosmological consequences of the modified
Friedmann equations when the entropy associated with the apparent
horizon, given by Barrow entropy, $S\sim A^{1+\delta/2}$, where
$0\leq\delta\leq1$, represents the amount of the
quantum-gravitational deformation of the horizon. We study
implications of this model in a flat Friedmann-Robertson-Walker
(FRW) universe with/without cosmological constant. Taking the
cosmological constant into account, this model can describe the
current accelerated expansion, although the transition from
deceleration phase to the acceleration phase takes place in the
lower redshifts. We investigate the evolution of the scale factor
and show that with increasing $\delta$, the value of the scale
factor increases as well. We also estimate the age of the universe
in Barrow cosmology which is smaller than the age of the universe
in standard cosmology.

\end{abstract}
 \maketitle

 \newpage
\section{Introduction\label{Intro}}
Inspired by the Covid-$19$ virus structure, recently Barrow
discussed \cite{Barrow} that quantum-gravitational effects may
deform the geometry of the black hole horizon leading to an
intricate, fractal features. He argued that the area law of the
black hole entropy get modified and is given by
\begin{eqnarray}\label{S}
S_{h}= \left(\frac{A}{A_{0}}\right)^{1+\delta/2},
\end{eqnarray}
where $A$ is the black hole horizon area and $A_0$ is the Planck
area. The exponent $\delta$ ranges as $0\leq\delta\leq1$ and
represents the amount of the quantum-gravitational deformation
effects. The area law is reproduced in case of $\delta=0$ and
$A_{0}\rightarrow 4G$. On the other hand, $\delta=1$ corresponds
to the most intricate and fractal structure of the horizon. In the
cosmological setup, the effects of Barrow entropy on the cosmic
evolution have been investigated from different viewpoints. For
example, modification of the area law leads to a new holographic
dark energy model based on Barrow entropy \cite{Emm1,Ana}. A
cosmological scenario based on Barrow entropy was proposed in
\cite{Emm2}, where it was shown that new extra terms that
constitute an effective dark energy sector appear in the Friedmann
equations. Although, it was argued in \cite{Emm2} that the
modified Friedmann equations based on Barrow entropy (\ref{S}) can
describe the thermal history of the universe from early
deceleration to the late time acceleration, with the dark-energy
epoch following the matter one, regardless of the presence of
cosmological constant $\Lambda$, nevertheless, it seems this
conclusion is only correct in the presence of cosmological
constant \cite{Emm2}. In other words, when $\Lambda=0$ the
equation of state (EoS) parameter of the dark sector is always
positive ($w_{DE}>0$) \cite{Emm2} (see Appendix). This implies
that the exponent $\delta$ in Barrow entropy cannot reproduce any
term which may play the role of dark energy \cite{Emm2}. On the
other hand, it was recently proven that Barrow entropy as well as
any other known entropy (Tsalis, Renyi, Kaniadakis, etc) is just
sub-case of generalized entropy expression introduced in
\cite{Odin1,Odin2}. Other studies on the cosmological consequences
of the Barrow entropy can be carried out in
\cite{Emm3,Abr1,Mam,Abr2,Bar2,Sri,Das,Sha,Pra,Odin3,Odin4}.

In the present work, we are going to investigate cosmological
implications of the modified Friedmann equations when the entropy
associated with the apparent horizon is given by the Barrow
entropy (\ref{S}). Our work differs from \cite{Emm2} in that the
author of \cite{Emm2} modifies the cosmological field equations in
such a way that leads to an extra component of energy in the
Friedmann equations.  In this approach the gravity side of the
Friedmann equation is not modified and the Barrow entropy acts as
an effective dark energy in the right hand side of the field
equations. However, our studies in the present work is based on
the modification of the geometry part (left hand side) of the
cosmological filed equations \cite{SheBFE} and keeping the energy
content of the universe in the form of ordinary matter and
radiation. This approach is well motivated and more physically
reasonable, since basically the entropy depends on the geometry of
spacetime (gravity part of the action). Any modification to the
entropy expression should affect directly the gravity side of the
field equations. In the Appendix of the present work, we compare
the results of \cite{Emm2} with the present work and clarify the
difference of our work with \cite{Emm2}. Throughout this paper we
set $\kappa_B=1=c=\hbar$, for simplicity.

This paper is structured as follows. In the next section, for
completeness, we briefly review the procedure of deriving the
modified Friedmann equations describing the evolution of the
universe, when the entropy associated with the apparent horizon is
in the form of Barrow entropy (\ref{S}). We shall see that the
cosmological constant can appear as a constant of integration in
the Friedmann equations. In section \ref{Cosmology}, we
investigate the cosmological consequences of the modified
Friedmann equations in the presence/absence of cosmological
constant. We also estimate the age of the universe in this
section. We finish with closing remarks in the last section.

\section{Modified Friedmann Equations based on Barrow entropy}\label{FIRST}
In this section, we briefly review the approach of constructing
the modified Friedmann equations based on Barrow entropy by using
the gravity-thermodynamics conjecture. We refer to \cite{SheBFE}
for details of calculations.

In the background of FRW universe, the line elements of the metric
is given by
\begin{equation}
ds^2=-dt^2+a^2(t)\left(\frac{dr^2}{1-kr^2}+r^2(d\theta^2+\sin^2\theta
d\phi^2)\right),
\end{equation}
where $a(t)$ is scale factor of the universe, $k = 0, \pm 1$ stand
for the curvature parameter, and $(t,r,\theta, \phi)$ are the
co-moving coordinates. We further assume $a_0=a(t=t_0)=1$, at the
present time. Assuming the apparent horizon as boundary of the
universe, the temperature associated with the horizon is given by
\cite{Cai1}
\begin{equation}\label{T}
T_h=-\frac{1}{2 \pi \tilde r_A}\left(1-\frac{\dot {\tilde
r}_A}{2H\tilde r_A}\right),
\end{equation}
where $\tilde{r}_A={1}/{\sqrt{H^2+k/a^2}}$ is the apparent horizon
radius \cite{Sheyem}. From the thermodynamical viewpoint the
apparent horizon is a suitable horizon consistent with first and
second law of thermodynamics \cite{wang1,wang2,Cai2,
Cai3,SheyCQ,sheyECFE}. We further assume the energy-momentum
tensor of the universe is $
T_{\mu\nu}=(\rho+p)u_{\mu}u_{\nu}+pg_{\mu\nu}$, where $\rho$ and
$p$ are the energy density and pressure, respectively. The
energy-momentum tensor is conserved, $\nabla_{\mu}T^{\mu\nu}=0$,
which results the continuity equation, $\dot{\rho}+3H(\rho+p)=0$
where $H=\dot{a}/a$ is the Hubble parameter. The work density
associated with the volume change of the expanding universe, is
also given by $W=(\rho-p)/2$ \cite{Hay2}. To employ the
gravity-thermodynamics conjecture, we propose the first law of
thermodynamics on the apparent horizon satisfies as
\begin{equation}\label{FL}
dE = T_h dS_h + WdV,
\end{equation}
where $E=\rho V$ is the total energy of the universe enclosed by
the apparent horizon, and $T_{h}$ and $S_{h}$ are, respectively,
the temperature and entropy associated with the apparent horizon.
Here $V=\frac{4\pi}{3}\tilde{r}_{A}^{3}$ is the volume enveloped
by a 3-dimensional sphere with the area of apparent horizon
$A=4\pi\tilde{r}_{A}^{2}$. Taking differential form of the total
matter and energy, we find $ dE=4\pi\tilde
 {r}_{A}^{2}\rho d\tilde {r}_{A}+\frac{4\pi}{3}\tilde{r}_{A}^{3}\dot{\rho} dt
$, which after combining with conservation equation, we arrive at
\begin{equation}
\label{dE2}
 dE=4\pi\tilde
 {r}_{A}^{2}\rho d\tilde {r}_{A}-4\pi H \tilde{r}_{A}^{3}(\rho+p) dt.
\end{equation}
Differentiating the Barrow entropy (\ref{S}), yields
\begin{eqnarray} \label{dS}
dS_h=(2+\delta)\left(\frac{4\pi}{A_{0}}\right)^{1+\delta/2}
 {\tilde
{r}_{A}}^{1+\delta} \dot{\tilde {r}}_{A} dt.
\end{eqnarray}
Finally, combining Eqs. (\ref{T}), (\ref{dE2}) and (\ref{dS}) with
the first law of thermodynamics (\ref{FL}), after some algebraic
calculations and using continuity relation, we arrive at
\begin{equation} \label{Fried2}
-\frac{2+\delta}{2\pi A_0
}\left(\frac{4\pi}{A_0}\right)^{\delta/2} \frac{d\tilde
{r}_{A}}{\tilde {r}_{A}^{3-\delta}}=
 \frac{d\rho}{3}.
\end{equation}
After integration, we find the first modified Friedmann equation
in Barrow cosmology,
\begin{equation} \label{Fried4}
\left(H^2+\frac{k}{a^2}\right)^{1-\delta/2} = \frac{8\pi G_{\rm
eff}}{3} \rho+\frac{\Lambda}{3},
\end{equation}
where $\Lambda$ is a constant of integration which can be
interpreted as the cosmological constant, and  $G_{\rm eff}$
stands for the effective Newtonian gravitational constant,
\begin{equation}\label{Geff}
G_{\rm eff}\equiv \frac{A_0}{4} \left(
\frac{2-\delta}{2+\delta}\right)\left(\frac{A_0}{4\pi
}\right)^{\delta/2}.
\end{equation}
If we define $\rho_{\Lambda}={\Lambda}/(8\pi G_{\rm eff})$, Eq.
(\ref{Fried4}), can be rewritten as
\begin{equation} \label{Fried5}
\left(H^2+\frac{k}{a^2}\right)^{1-\delta/2} = \frac{8\pi G_{\rm
eff}}{3}(\rho+\rho_{\Lambda}).
\end{equation}
In this way, we derive the modified Friedmann equation by starting
from the first law of thermodynamics, and assuming the entropy
associated with the apparent horizon has the form (\ref{S}). When
$\delta=0$, the area law of entropy is restored and
$A_{0}\rightarrow4G$. In this case, $G_{\rm eff}\rightarrow G$,
and Eq. (\ref{Fried4}) reduces to the standard Friedmann equation
in General Relativity.

To get the second Friedmann equation, we can combine the
continuity equation with the first Friedmann equation
(\ref{Fried4}). It is a matter of calculations to show that
\cite{SheBFE}
\begin{eqnarray}
&&(2-\delta)\frac{\ddot{a}}{a}
\left(H^2+\frac{k}{a^2}\right)^{-\delta/2}+(1+\delta)\left(H^2+\frac{k}{a^2}\right)^{1-\delta/2}
\nonumber
\\
&&=-8\pi G_{\rm eff}(p+p_{\Lambda}),\label{2Fried3}
\end{eqnarray}
where $p_{\Lambda}=-{\Lambda}/(8\pi G_{\rm eff})$. This is the
second modified Friedmann equation governing the evolution of the
universe based on Barrow entropy. In the limiting case where
$\delta=0$ ($G_{\rm eff}\rightarrow G$), Eq. (\ref{2Fried3})
reduces to the second Friedmann equation in standard cosmology.
Combining Eqs. (\ref{Fried4}) and (\ref{2Fried3}), yields
\begin{eqnarray}
&&(2-\delta)\frac{\ddot{a}}{a}\left(H^2+\frac{k}{a^2}\right)^{-\delta/2}\nonumber\\&&
=-\frac{8\pi
G_{\rm eff}}{3} \left[3(p+p_{\Lambda})+(\delta+1)(\rho+\rho_{\Lambda})\right]\nonumber\\
&&=-\frac{8\pi G_{\rm eff}}{3} (\rho+\rho_{\Lambda}) \left(3w_{\rm
tot}+\delta+1\right), \label{2Frie5}
\end{eqnarray}
where $w_{\rm tot}$ is the EoS parameter of the total energy and
matter, defined as
\begin{eqnarray}
&&w_{\rm tot}=\frac{p+p_{\Lambda}}{\rho+\rho_{\Lambda}}.
\end{eqnarray}
From Eq. (\ref{2Frie5}), one may notice that the condition for the
acceleration of the cosmic expansion ($\ddot{a}>0$), yields
\begin{eqnarray}
 1+\delta +3 w_{\rm
tot}<0  \  \     \longrightarrow   \  \  w_{\rm tot}<
-\frac{1+\delta}{3}.          \label{w1}
\end{eqnarray}
For $\delta=0$, we have $w_{\rm tot}<-1/3$, while for $\delta=1$
we find $w_{\rm tot}<-2/3$. The former is the standard cosmology,
while the later represents the most intricate and fractal
structure of the horizon. As we shall see in the next section, our
model can reproduce the accelerated expansion provided we take the
cosmological constant into account. This is, perhaps, the main
difference between Barrow and Tsallis cosmology \cite{SheTs},
despite the origin of the corrections in entropy which are
completely different in these two cases  \cite{Tsa}. It was argued
that choosing the non-extensive parameter in Tsallis cosmology as
$\beta<1/2$, leads to an accelerated universe, without invoking
any kind of dark energy (cosmological constant) \cite{SheTs} (see
also \cite{Matin,Maj} for dark energy models based on Tsallis
entropy). However, in Barrow cosmology, we observe that in order
to have $\ddot {a}>0$, the EoS parameter should be always negative
in the allowed range of $\delta$, requiring an additional
component of dark energy/cosmological constant to reproduce an
accelerated universe \cite{Emm2}.

Given the modified Friedmann equations (\ref{Fried5}) and
(\ref{2Fried3}) at hand, in the next section, we investigate the
cosmological implications of this model.
\section{Modified Cosmology\label{Cosmology}}
In this section we are going to investigate cosmological
consequences of the modified Friedmann equations given in Eqs.
(\ref{Fried5}) and (\ref{2Fried3}). For simplicity, we focus on
the flat universe ($k=0$), although the study can be carried out
for $k=\pm1$.

\subsection{The case $\Lambda=0$}
Let us first consider  the case where the cosmological constant is
zero ($\Lambda=0$) and the universe is dominated with pressureless
matter. Integrating, the continuity equation $\dot{\rho_{
m}}(t)+3H\rho_{ m}(t)=0$, immediately yields
\begin{eqnarray} \label{rhom}
\rho_{ m}(t)\propto a^{-3} \   \Rightarrow \  \rho_{m}(t)=C_1
a^{-3},
\end{eqnarray}
where $C_1$ is a constant of proportionality. In order to derive
the evolution of the scale factor, we insert $\rho_{m}$ from
(\ref{rhom}) in the modified Friedmann equation in Barrow
cosmology (\ref{Fried5}). We find
\begin{eqnarray}
\left(\frac{\dot{a}}{a}\right)^{2-\delta}= \frac{8 \pi G_{\rm
eff}}{3} C_1 a^{-3},
\end{eqnarray}
which can be rewritten as
\begin{eqnarray} \label{adot}
\left(\frac{d{a}}{dt}\right)^{2-\delta}= C_2 a^{-1-\delta},
\end{eqnarray}
where the constant $C_2$ is defined
\begin{eqnarray}
C_2\equiv\frac{8 \pi G_{\rm eff} C_1}{3}.
\end{eqnarray}
One can easily integrate Eq. (\ref{adot}), which has the solution
\begin{eqnarray}
a(t)=  C_3 t^{(2-\delta)/3},
\end{eqnarray}
where
\begin{eqnarray}
C_3\equiv \left[\frac{3}{2-\delta} C_2
^{1/(2-\delta)}\right]^{(2-\delta)/3} .
\end{eqnarray}
When $\delta=0$, we have
\begin{eqnarray}
a(t)= \left[\frac{3}{2}\sqrt{C_2}\right]^{2/3} t^{2/3},
\end{eqnarray}
which is the well-known result of standard cosmology.  The second
time derivative of the scale factor is given by
\begin{eqnarray}
\ddot{a}(t)= - \frac{C_3}{9}(1+\delta)(2-\delta) \
t^{-(4+\delta)/3},
\end{eqnarray}
which is always negative ($\ddot{a}<0$) during the evolution of
the universe in the allowed range of $\delta$. Thus, in Barrow
cosmology, one should take into account a dark energy/cosmological
constant component to explain the acceleration of the cosmic
expansion. This is consistent with the argument given in
\cite{Emm2}.

The evolution of the energy density, the Hubble and the
deceleration parameters are calculated as
\begin{eqnarray}
\rho_{m}(t)& \propto & \frac{1}{t^{2-\delta}},\\
H(t)&=& \frac{\dot{a}}{a}=\frac{2-\delta}{3 t}, \label{Hubble}\\
q(t)&=& -1-\frac{\dot{H}}{H^2}=\frac{1+\delta}{2-\delta}.
\end{eqnarray}
Again, all above parameters reduce to those of standard cosmology
for $\delta=0$. Looking at the deceleration parameter indicates
$q>0$, implying a decelerated universe in Barrow cosmology filled
with pressureless matter.

Now we consider a universe filled with radiation. This case makes
only sense at the early stage of the universe where the radiation
was dominated. Since our Universe is expanding, the proper momenta
of freely moving particles decreases as $P(t)\sim1/a(t)$. This
implies that the random velocities of particles seen today should
have been large in the past when the scale factor was much smaller
than its present value \cite{GRbook}. As a result, the
pressureless approximation is break down in the early universe.
Our aim here is to obtain the evolution of the Universe in the
framework of Barrow cosmology, when the energy content of the
universe is composed of highly relativistic gas (radiation) with
EoS $p_{ r}=\rho_{ r}/3$. In this case from the continuity
equation, $\dot{\rho_{ r}}(t)+4H\rho_{ r}(t)=0$, we can get
\begin{eqnarray} \label{rhor}
\rho_{ r}(t)\propto a^{-4} \   \Rightarrow \  \rho_{ r}(t)=B_1
a^{-4},
\end{eqnarray}
where $B_1$ is an integration constant. Combining $\rho_{\rm
r}(t)$ given in (\ref{rhor}) with the first Friedmann equation
(\ref{Fried5}) for $k=0=\Lambda$, one gets
\begin{eqnarray}
\left(\frac{\dot{a}}{a}\right)^{2-\delta}= \frac{8 \pi G_{\rm eff}
}{3} B_1 a^{-4},
\end{eqnarray}
which can be rewritten as
\begin{eqnarray} \label{adotr}
\left(\frac{d{a}}{dt}\right)^{2-\delta}= B_2 a^{-2-\delta},
\end{eqnarray}
where we have defined
\begin{eqnarray}
B_2\equiv\frac{8 \pi G_{\rm eff}}{3}B_1.
\end{eqnarray}
Solving Eq. (\ref{adotr}) for scale factor, we arrive at
\begin{eqnarray}\label{ar}
a(t)= B_3 \ t^{(2-\delta)/4}.
\end{eqnarray}
where
\begin{eqnarray}
B_3=\left[\frac{4}{2-\delta}B_2^{1/(2-\delta)}\right]^{(2-\delta)/4}.
\end{eqnarray}
The standard cosmology is deduced by setting  $\delta=0$, yielding
\begin{eqnarray}
a(t)= \sqrt{2}B_2^{1/4} \ t^{1/2}.
\end{eqnarray}
Taking the second time derivative of the scale factor (\ref{ar})
leads to
\begin{eqnarray}
\ddot{a}(t)= - B_3 \left(\frac{4-\delta^2}{16}\right)
t^{-(6+\delta)/4}<0.
\end{eqnarray}
which is an expected result for radiation dominated era, where the
universe was undergoing a decelerated phase ($\ddot{a}(t)<0$).
Now, we calculate the energy density, the Hubble and the
deceleration parameters in the radiation dominated era. We find
\begin{eqnarray}
\rho_{r}(t)& \propto & \frac{1}{t^{2-\delta}},\\
H(t)&=& \frac{\dot{a}}{a}=\frac{2-\delta}{4 t},\\
q(t)&=& -1-\frac{\dot{H}}{H^2}=\frac{2+\delta}{2-\delta}>0,
\end{eqnarray}
which again confirm that $q>0$ ($\ddot{a}(t)<0$). This implies
that in the early stage of the universe where the relativistic
particles have been dominated, our Universe has been in a
decelerated phase. In summary, in the presence of radiation and
pressureless matter, Barrow cosmology cannot explain the cosmic
phase transition from a deceleration phase to an acceleration
phase during the history of the universe, unless the dark
energy/cosmological constant is taken into account \cite{Emm2}.
This is in contrast to the Tsallis cosmology \cite{SheTs} where by
suitably choice of the nonextensive parameter, it is quite
possible to explain the history of the universe from a
deceleration to acceleration phase without invoking any kind of
dark energy/cosmological constant.
\subsection{The case with $\Lambda\neq0$}
In this case, we define the density parameters as
\begin{eqnarray} \label{DP}
\Omega_m=\frac{\rho_m}{\rho_{c}}, \  \  \
\Omega_{\Lambda}=\frac{\rho_{\Lambda}}{\rho_{c}}, \  \   \
\rho_{c}=\frac{3H^{2-\delta}}{8\pi G_{\rm eff}}. \  \
\end{eqnarray}
Therefore, in terms of the density parameters, the first Friedmann
equation (\ref{Fried5}) can be written as
\begin{eqnarray}
\Omega_m+ \Omega_{\Lambda}=(1+\Omega_k)^{1-\delta/2},
\end{eqnarray}
where, as usual, the curvature density parameter is given by
$\Omega_k=k/(a^2H^2)$. We consider a flat universe filled with
pressureless matter ($p=p_m=0$) and cosmological constant, and
hence
\begin{eqnarray}
\Omega_m+ \Omega_{\Lambda}=1.
\end{eqnarray}
\begin{figure}[ht]
\includegraphics[scale=0.8]{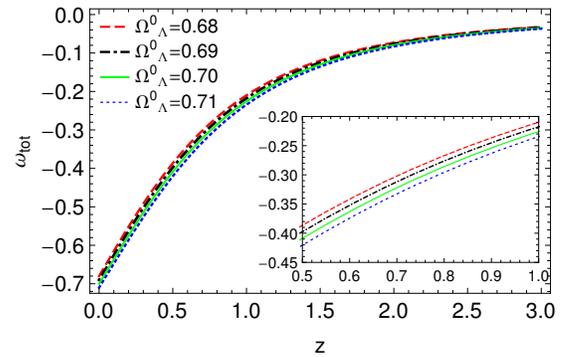}
\caption{Evolution of $w_{\rm tot}$ as a function of redshift
parameter $z$ in modified Barrow cosmology for different values of
$\Omega^{0}_{\Lambda}$.}\label{Fig1}
\end{figure}
The total EoS parameter can be written
\begin{eqnarray}
&&w_{\rm
tot}=\frac{p_m+p_{\Lambda}}{\rho_m+\rho_{\Lambda}}=\frac{p_{\Lambda}}{\rho_m+\rho_{\Lambda}}=\frac{p_{\Lambda}/\rho_{\Lambda}}{\rho_m/\rho_{\Lambda}+1},\nonumber \\
&&=-\frac{1}{\rho_m/\rho_{\Lambda}+1},
\end{eqnarray}
where $p_{\Lambda}/\rho_{\Lambda}=-1$, according to the definition
of $p_{\Lambda}$ and $\rho_{\Lambda}$. Taking into account the
fact that $\rho_{\Lambda}=\rho^{0}_{\Lambda}$, and
$\rho_m=\rho^{0}_m (1+z)^{3}$, and using the definition of density
parameters in Eq. (\ref{DP}), we obtain
\begin{eqnarray}
w_{\rm tot} (z)=-\frac{\Omega^{0}_{\Lambda}}{\Omega^{0}_m
(1+z)^3+\Omega^{0}_{\Lambda}}.
\end{eqnarray}
If we take $\Omega^{0}_{\Lambda} \simeq 0.7 $ and
$\Omega^{0}_m\simeq 0.3$, we have
\begin{eqnarray}
&&w_{\rm tot} (z)=-\frac{0.7}{0.7+0.3 (1+z)^3}.
\end{eqnarray}
At the present time where $z\rightarrow 0$, we have $w_{\rm
tot}=-0.7$, while at the early universe where $z\rightarrow
\infty$, we get $w_{\rm tot}=0$. This implies that at the early
stages, the universe undergoes a decelerated phase while at the
late time it experiences an accelerated phase. The behaviour of
the total EoS parameter in term of the redshift is plotted in Fig.
1. This figure shows that $w_{\rm tot}$ decreases with increasing
$\Omega^{0}_{\Lambda}$, which is an expected result.

We can also obtain the deceleration parameter defined as
$q=-1-\dot{H}/{H^2}$. It is a matter of calculations to show that
 \begin{eqnarray}
&&q(z)=-1+\frac{3}{2-\delta}(1-\Omega^{0}_{\Lambda})(1+z)^3\nonumber\\
&&\times\left[{\Omega^{0}_{\Lambda}+(1-\Omega^{0}_{\Lambda})(1+z)^3}\right]^{-1}.
\end{eqnarray}
The behaviour of $q$ versus $z$ are plotted in Figs. 2 and 3. In
Fig. 2, we keep $\delta=0.4$ and investigate the effects of
$\Omega^{0}_{\Lambda}$ on $q(z)$. We observe that $q$ is not
sensitive to the present values of $\Omega^{0}_{\Lambda}$, while
in Fig. 3, we see that $q$ is very sensitive to the Barrow
exponent $\delta$. From Fig. 3, we see that with increasing
$\delta$, the transition from deceleration phase ($q>0$) to the
acceleration phase ($q<0$) takes place at lower redshifts. Indeed,
the best consistency with observation for the phase transition
happens for $\delta=0$ at $z\approx 0.63$.
\begin{figure}[ht]
\includegraphics[scale=0.8]{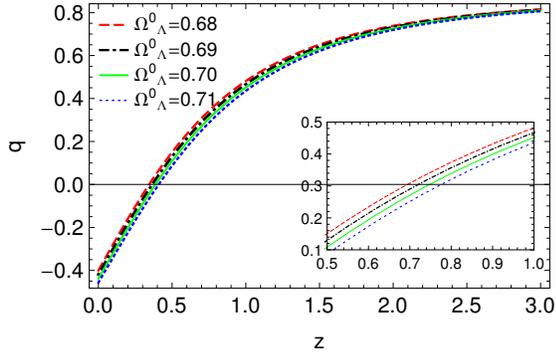}
\caption{Evolution of $q$ as a function of redshift parameter $z$
in modified Barrow cosmology for $\delta=0.4$ and different values
of $\Omega^{0}_{\Lambda}$.}\label{Fig2}
\end{figure}
\begin{figure}[ht]
\includegraphics[scale=0.8]{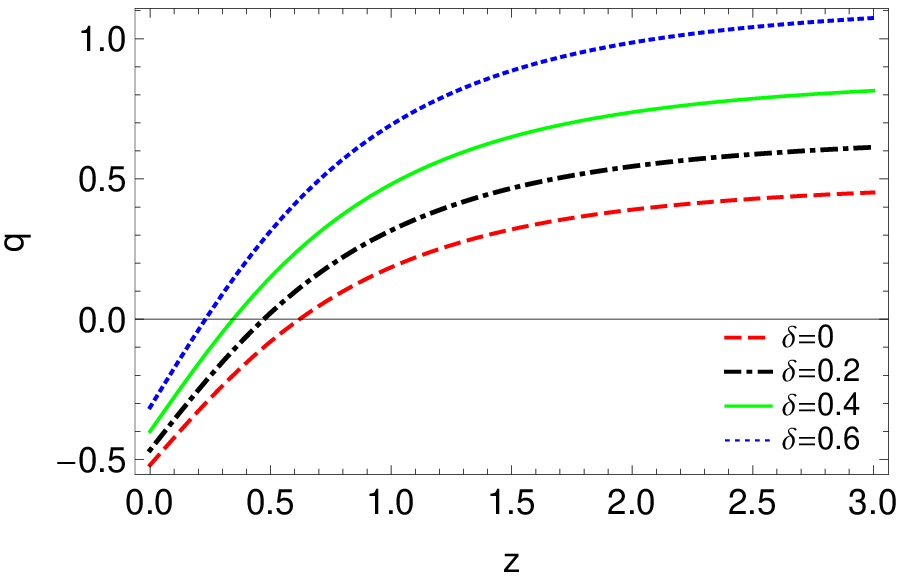}
\caption{Evolution of $q$ as a function of redshift parameter $z$
in modified Barrow cosmology for $\Omega^{0}_{\Lambda}=0.68$ and
different values of $\delta$.}\label{Fig3}
\end{figure}
\begin{figure}[ht]
\includegraphics[scale=0.6]{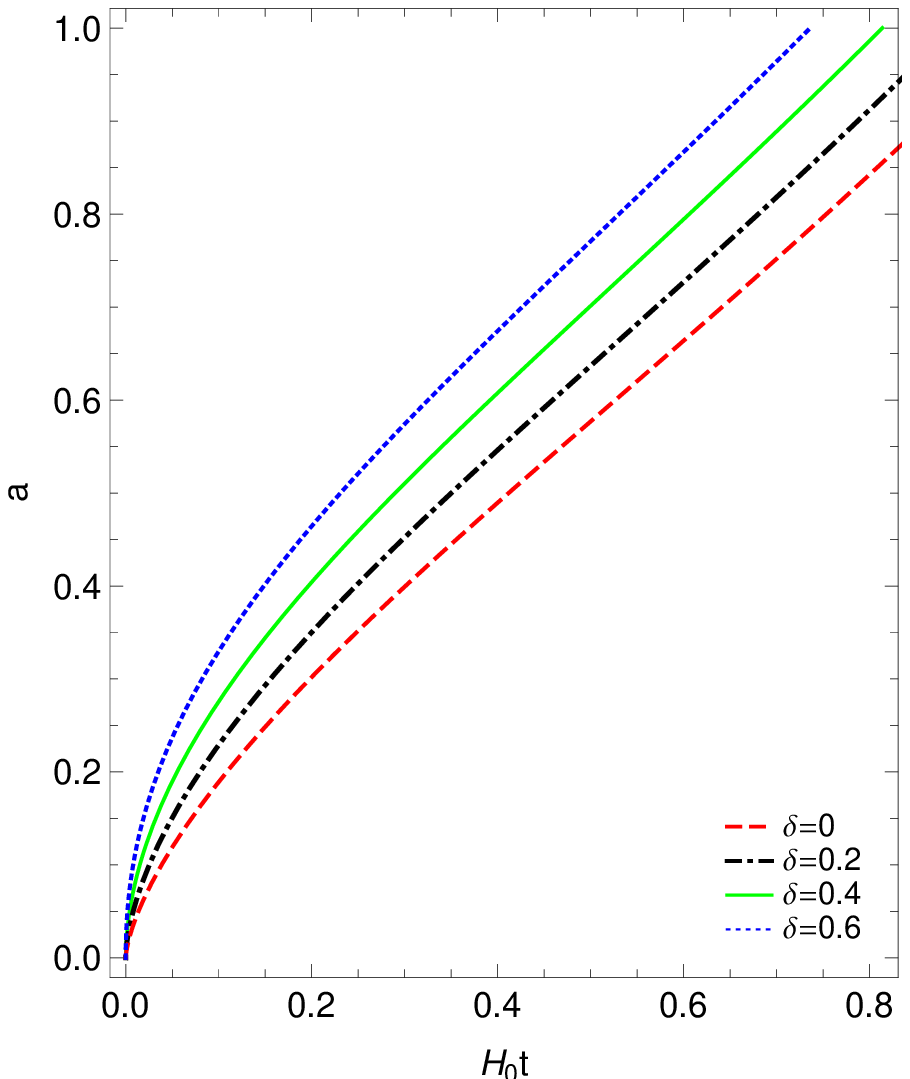}
\caption{Evolution of scale factor $a$ versus $H_{0}t$ in modified
Barrow cosmology for $\Omega^{0}_{\Lambda}=0.70$ and different
values of $\delta$.}\label{Fig4}
\end{figure}
\begin{figure}[ht]
\includegraphics[scale=0.6]{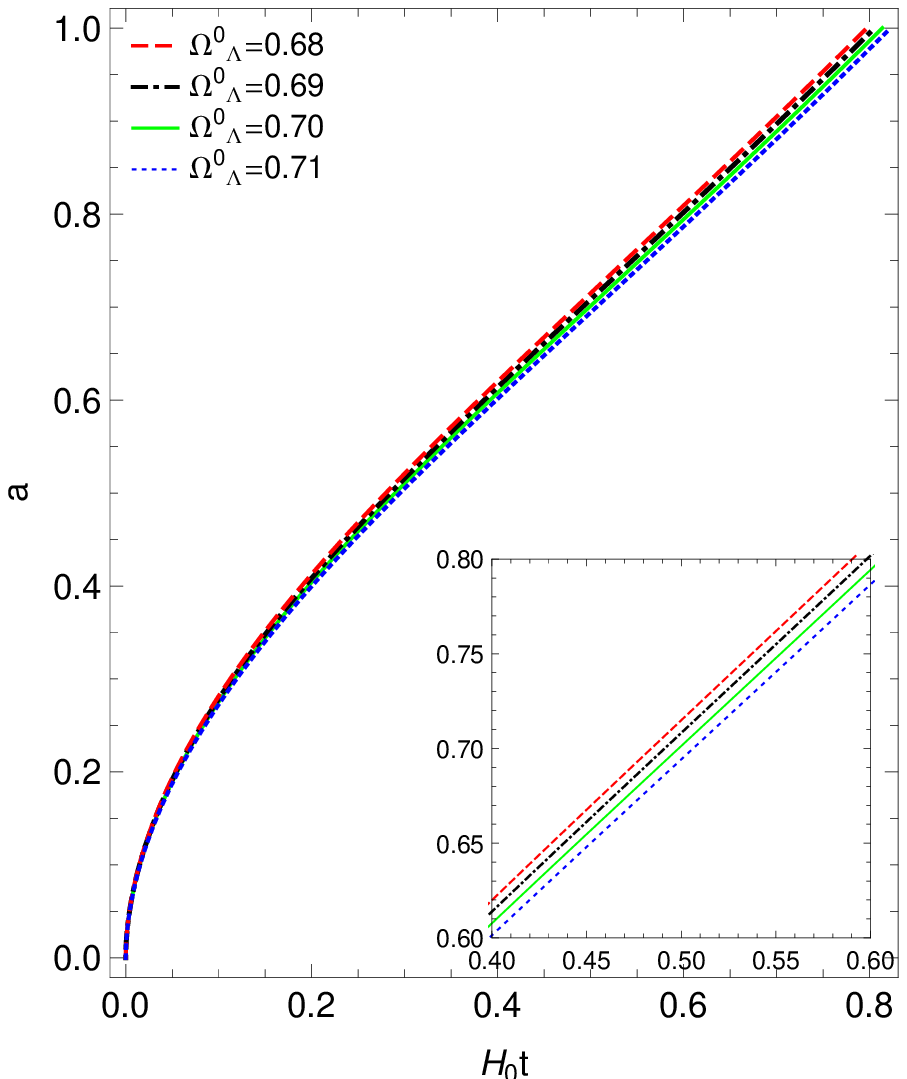}
\caption{Evolution of scale factor $a$ versus $H_{0}t$ in modified
Barrow cosmology for $\delta=0.4$ and different values of
$\Omega^{0}_{\Lambda}$.}\label{Fig5}
\end{figure}

Next, we investigate the scale factor of the universe. The first
Friedmann equation (\ref{Fried5}) for flat universe ($k=0$) can be
rewritten as
\begin{eqnarray}\label{H}
&&H^{2-\delta}=H^{2-\delta}_{0}\left[\Omega^{0}_m a^{-3}+
\Omega^{0}_{\Lambda}\right],\nonumber \\ \Rightarrow
&&\frac{da}{dt}=H_{0} a\left[\Omega^{0}_m a^{-3}+
\Omega^{0}_{\Lambda } \right]^{1/(2-\delta)},
\end{eqnarray}
where $H_0=H(t=t_0)$ is the Hubble parameter at the present time
$t_0$. Integrating (\ref{H}), yields
\begin{eqnarray}
&&H_{0} t=\int{a^{-1}  \left[(1-\Omega^{0}_{\Lambda }) a^{-3}+
\Omega^{0}_{\Lambda } \right]^{1/(\delta-2)}da}.
\end{eqnarray}
Let us look at the above relation for the special case where
$\Omega^{0}_{\Lambda}\approx 1$ and $\Omega^{0}_m\approx 0$. In
this case, we have $H_0 t\sim \ln a,$ and hence $a(t)\sim
\exp(H_{0} t)$, which describes a de-Sitter universe, independent
of the value of $\delta$. For $\Omega^{0}_{\Lambda} \simeq 0.7 $
and $\Omega^{0}_m\simeq 0.3$, we have
\begin{eqnarray} \label{H0t}
&&H_{0} t=\int{a^{-1}  \left[0.3 a^{-3}+ 0.7
\right]^{1/(\delta-2)}da}.
\end{eqnarray}
In principle, one can solve the Eq. (\ref{H0t}) to obtain the
scale factor of the universe for different values of $\delta$.
Indeed, for a given value of $\delta$, it is better to plot the
scale factor $a$ versus $H_0 t$ which are shown in Figs. 4-5. From
Fig. 4, we see that at each time, the scale factor increases with
increasing $\delta$. Therefore, in modified Barrow cosmology the
radius of the universe increases comparing to the standard
cosmology.
\subsection{Age of the universe\label{age}}
Given the Hubble parameter at hand, we can estimate the age of the
universe at the present time ($t=t_0$). In the absence of
cosmological constant ( $\Lambda=0$) and for the matter dominated
universe, from (\ref{Hubble}) we can estimate the age of the
universe in modified Barrow cosmology as
\begin{equation}\label{age}
t_0|_B=\frac{2-\delta}{3H_0}=\frac{2-\delta}{2}t_0|_S ,
\end{equation}
where $H_0=H(t_0)$ is the Hubble constant and $t_0|_S={2}/(3H_0)$
is the age of the universe in standard cosmology with $\Lambda=0$.
Thus, compared to the standard cosmology, the age of the universe
decreases by factor ${(2-\delta)}/2<1$. This implies that in
modified cosmology based on Barrow entropy, the age problem cannot
be alleviated. This is in contrast to the Tsallis cosmology
\cite{SheTs}, where the age problem can be alleviated provided one
take the non-extensive parameter $\beta<1/2$ \cite{SheTs}. When we
take the cosmological constant into account, the age of the
universe can be obtained through relation
\begin{eqnarray}
t_0-t=\int_0^{z}{\frac{dz}{H(z)(1+z)}},
\end{eqnarray}
where $a=(1+z)^{-1}$. Substituting $H(z)$ from Eq. (\ref{H}), we
arrive at
\begin{eqnarray}
t_0-t=\frac{1}{H_0}\int_0^{z}{\frac{dz}{(1+z)\left[\Omega^{0}_m
(1+z)^{3}+ \Omega^{0}_{\Lambda } \right]^{1/(2-\delta)}}}.
\end{eqnarray}
In order to obtain the age of the universe, we must take the limit
$t\rightarrow0$ when $z\rightarrow \infty$. Rewriting the above
expression in terms of $x=(1+z)^{-1}$, we have
\begin{eqnarray}
t_0=\frac{1}{H_0}\int_0^{1}{\frac{dx}{x\left[\Omega^{0}_m x^{-3}+
\Omega^{0}_{\Lambda } \right]^{1/(2-\delta)}}}.
\end{eqnarray}
We observe that the age of the Universe in this model depends on
$\delta$. In table I we present the values of $H_{0} t_{0}$ for
different $\delta$ and fixed value of $\Omega^{0}_{\Lambda }$. We
observe that with increasing the value of $\delta$, the age of the
universe decreases. Besides, from table II, we see that for a
fixed value of $\delta$, the age of the universe increases with
increasing $\Omega^{0}_{\Lambda }$.
\begin{table}[t]
\begin{center}
\begin{tabular}{c|c|c|c|c|}
\hline
\multicolumn{1}{|c|}{$\delta$} & $0$ & $0.2$ & $0.4$ & $0.6$ \\
\hline
\multicolumn{1}{|c|}{$H_{0} t_{0}$} & $0.9468$ &  $0.8725$ & $0.7970$ &  $0.7199$ \\
\hline
\end{tabular}
\caption{Numerical results for $H_0 t_0$ in modified Barrow
cosmology for  $\Omega^{0}_{\Lambda}=0.68$ and different values of
$\delta$.}\label{table1}
\end{center}
\end{table}
\begin{table}[t]
\begin{center}
\begin{tabular}{c|c|c|c|c|}
\hline
\multicolumn{1}{|c|}{$\Omega^{0}_{\Lambda}$} & $0.68$ & $0.69$ & $0.70$ & $0.71$ \\
\hline
\multicolumn{1}{|c|}{$H_{0} t_{0}$} & $0.7970$ &  $0.8052$ & $0.8136$ &  $0.8224$ \\
\hline
\end{tabular}
\caption{Numerical results for $H_0 t_0$ in modified Barrow
cosmology for $\delta=0.4$ and different values of
$\Omega^{0}_{\Lambda}$.}\label{table1}
\end{center}
\end{table}
\section{Closing remarks \label{Con}}
The cosmological field equations govern the evolution of the
universe get modified, due to the quantum-gravitational
deformation effects of the apparent horizon. In this work, we have
investigated the cosmological consequences of the modified
Friedmann equations when the entropy associated with the apparent
horizon is in the form of Barrow entropy (\ref{S}). We showed that
in the presence of cosmological constant, this model can explain
the current accelerated universe, although the transition from
decelerated phase ($q>0$) to the accelerated phase ($q<0$) takes
place in the lower redshifts, compared to the standard cosmology.
We obtained the scale factor, Hubble parameter and deceleration
parameter in the presence/absence of cosmological constant which
depend on the Barrow exponent $\delta$. We have also estimated the
age of the universe in this model and observed that comparing to
the standard cosmology ($\delta=0$), the age of the universe
decreases with increasing $\delta$. On the other hand for each
value of $\delta$, the age of the universe increases with
increasing $\Omega^{0}_{\Lambda}$.

In conclusion, the main result obtained in the present work is
that, in modified Barrow cosmology without cosmological constant
($\Lambda=0$), one cannot deduce a thermal history of the universe
compatible with observations and one needs to take into account
the cosmological constant to reproduce an accelerated universe. In
other words, the exponent $\delta$ in Barrow entropy cannot
reproduce any term in the dynamical cosmological equations which
may act as the dark energy sector.

It is also interesting to study the profile of the growth of
density perturbation in the context of Barrow cosmology. The
details of investigations on the density perturbation as well as
the gravitational collapse in the background of Barrow cosmology
will be addressed in the future projects.
 \section*{Appendix: Comparing with ``Modified cosmology through spacetime
thermodynamics and Barrow horizon entropy'' \cite{Emm2} }

Here we review the modified cosmology based on Barrow entropy
discussed in \cite{Emm2} and compare the present work with it in
more details. Taking into account the entropy associated with the
apparent horizon in the form of Barrow entropy given in Eq.
(\ref{S}), the author of \cite{Emm2} applies the first law of
thermodynamics, $-dE=TdS$, on the apparent horizon and derives the
modified Friedmann equations in Barrow cosmology. Here $-dE$ is
the energy flux crossing the apparent horizon within an
infinitesimal internal of time $dt$. While in the present work, we
take the first law of thermodynamics as $dE=TdS+WdV$, where $dE$
is now the change in the energy inside the apparent horizon.
Besides, in \cite{Emm2} the apparent horizon radius $\tilde r_A$
has been assumed to be fixed. Thus, the temperature of apparent
horizon can be approximated to $T=1/(2\pi \tilde r_A)$ and there
is no the term of volume change in it. But, here, we have used the
matter energy $E=\rho V$ inside the apparent horizon and the
apparent horizon radius changes with time. This is the reason why
we have included the term $WdV$ in the first law (\ref{FL}).

For a flat universe the modified Friedmann equations, based on
Barrow entropy, derived in \cite{Emm2}, are given by
\begin{eqnarray}
\label{FR1}
&&H^2=\frac{8\pi G}{3}\left(\rho_m+\rho_{DE}\right)\\
&&\dot{H}=-4\pi G \left(\rho_m+p_m+\rho_{DE}+p_{DE}\right),
\label{FR2}
\end{eqnarray}
where the energy density and pressure of the effective dark energy
are defined as \cite{Emm2}
 \begin{eqnarray}\label{FrE1}
&& \rho_{DE}=\frac{3}{8\pi G} \left\{
\frac{\Lambda}{3}+H^2\left[1-\frac{ \beta (\delta+2)}{2-\delta}
H^{-\delta} \right] \right\},\\
&& p_{DE}= -\frac{1}{8\pi G}\Bigg{\{} \Lambda
+2\dot{H}\left[1-\beta\left(1+\frac{\delta}{2}\right) H^{-\delta}
\right]\nonumber\\
&&+3H^2\left[1-\frac{\beta(2+\delta)}{2-\delta}H^{-\delta} \right]
\Bigg{\}}, \label{FrE2}
\end{eqnarray}
where $\beta$ and $\Lambda$ are constants \cite{Emm2}. It is clear
that the Friedmann equations (\ref{FR1}) and (\ref{FR2}) differ
from the modified Friedmann equations we derived in Eqs.
(\ref{Fried5}) and (\ref{2Fried3}) of the present work. According
to approach \cite{Emm2} the gravity side of the Friedmann equation
is not modified and the Barrow entropy acts as an effective dark
energy sector in the right hand side of the field equations.
However, in our work, we keep the energy content of the universe
in the form of ordinary matter and radiation and the left hand
side of the Friedmann equations get modified due to the correction
to the entropy. We believe this is more reasonable, since
basically the entropy expression depends on the geometry of
spacetime (gravity part of the action). Any modification to the
entropy expression should influence the gravity side (left hand
side) of the field equations.

Considering the effective dark energy in the Friedmann equations
(\ref{FR1}) and (\ref{FR2}), one can define the effective EoS
parameter as
\begin{eqnarray}
w_{DE}\equiv\frac{p_{DE}}{\rho_{DE}}=-1- \frac{
2\dot{H}\left[1-\beta\left(1+\frac{\delta}{2}\right) H^{-\delta}
\right]
 }{\Lambda+3H^2\left[1-\frac{\beta(2+\delta)}{2-\delta}H^{-\delta}
\right]} \label{FRWwDE}.
\end{eqnarray}
The cosmological implications of the modified Friedmann equations
(\ref{FR1}) and (\ref{FR2}) were studied in \cite{Emm2}. It was
shown that for $\Lambda=0$ the EoS parameter of the dark sector
behaves as $w_{DE}=\delta/(2-\delta) \Omega_{DE}^{-1} $ which is
always positive ($w_{DE}>0$) in the allowed range of exponent
$0\leq\delta\leq1$. This means that exponent $\delta$ in Barrow
entropy cannot reproduce any term which may play the role of dark
energy. In other words, in order to obtain the thermal history of
the universe in agreement with observations and reproduce the late
time acceleration in the context of Barrow cosmology, one needs to
consider an additional dark energy (cosmological constant) in the
energy content of the universe. This is consistent with the
results obtained in the present work.

\acknowledgments{I am grateful to the referee for valuable and
constructive comments which helped me improve the paper
significantly. I also thank Dr. Mahya Mohammadi for helpful
discussion and valuables comments.}


\end{document}